\documentclass[11pt,a4paper]{article}
\usepackage{jcappub}
\usepackage{graphicx}% Include figure files
\usepackage{dcolumn}% Align table columns on decimal point
\usepackage{bm}% bold math
\usepackage{hepunits}
\usepackage{amsmath,amssymb,amsfonts}
\usepackage{mathrsfs}
\usepackage{color}
\usepackage{hyperref}
\usepackage{subfigure}
\usepackage{epsfig,graphicx,psfrag,axodraw,bm,amssymb}

\long\def\symbolfootnote[#1]#2{\begingroup%
\def\thefootnote{\fnsymbol{footnote}}\footnote[#1]{#2}\endgroup}
\def\lsim{\mathrel{\rlap{\lower4pt\hbox{\hskip1pt$\sim$}}
    \raise1pt\hbox{$<$}}}
\newcommand{\cog}{CoGeNT }
\newcommand{\EE}{\textbf{E}}
\newcommand{\EEc}{\textbf{E$^c$}}
\newcommand{\PP}{\textbf{P}}
\newcommand{\PPc}{\textbf{P$^c$}}
\newcommand{\be}{\begin{eqnarray}}
\newcommand{\ee}{\end{eqnarray}}
\newcommand{\uone}{$U(1)$}

  \def\lsim{\mathrel{\rlap{\lower4pt\hbox{\hskip2pt$\sim$}}
    \raise 13 pt\hbox{$<$}}}
    
\def\gsim{\mathrel{\rlap{\lower4pt\hbox{\hskip0.2pt$\sim$}}
    \raise 2pt\hbox{$>$}}}

\newcommand{\ba}{\begin{array}}  
\newcommand{\ea}{\end{array}}

\newcommand{\varphie}{\varphi_{\textrm{\textbf{e}}}}
\newcommand{\varphip}{\varphi_{\textrm{\textbf{p}}}}

\long\def\symbolfootnote[#1]#2{\begingroup%
\def\thefootnote{\fnsymbol{footnote}}\footnote[#1]{#2}\endgroup}

\begin{document}

{\scriptsize FERMILAB-PUB-11-207-T} \\  
{\scriptsize MIT-CTP 4249} \\

%\preprint{APS/123-QED}

\title{\LARGE{Dark Atoms: Asymmetry and Direct Detection}}

\author[a]{David E. Kaplan,}
\author[a,b]{Gordan Z. Krnjaic}
\author[c]{Keith R. Rehermann}
\author[d]{Christopher M. Wells}

\affiliation[a]{Department of Physics and Astronomy, The Johns Hopkins University\\
3400 N. Charles Street, Baltimore, MD}
\affiliation[b]{Theoretical Physics Group, Fermi National Accelerator Laboratory \\  Batavia, IL}
\affiliation[c]{Center for Theoretical Physics, MIT \\ 77 Mass Ave., Cambridge, MA}
\affiliation[d]{Department of Physics, Houghton College \\ 1 Willard Avenue, Houghton, NY}

\emailAdd{dkaplan@pha.jhu.edu}
\emailAdd{gordan@pha.jhu.edu}
\emailAdd{krmann@mit.edu}
\emailAdd{christopher.wells@houghton.edu}
%\author{David E. Kaplan}
%\email[]{dkaplan@pha.jhu.edu}
%\affiliation{Department of Physics and Astronomy \\ The Johns Hopkins University
%3200 N. Charles Street\\ Baltimore, MD}
%
%\author{Gordan Z. Krnjaic}
%\email[]{gordan@pha.jhu.edu}
%\affiliation{Department of Physics and Astronomy \\ The Johns Hopkins University
%3400 N. Charles Street\\ Baltimore, MD\\}
%\affiliation{\\Theoretical Physics Group \\ Fermi National Accelerator Laboratory \\  Batavia, IL}
%
%\author{Keith R. Rehermann}
%\email[]{krmann@mit.edu}
%\affiliation{Center for Theoretical Physics \\ MIT \\ 77 Mass Ave. \\ Cambridge, MA}
%
%\author{Christopher M. Wells}
%\email[]{christopher.wells@houghton.edu}
%\affiliation{Department of Physics \\ Houghton College \\ 1 Willard Avenue \\ Houghton, NY}

%\author{Usual Suspects}
%\email[]{blagh@pha.jhu.edu}
%\affiliation{Department of Physics and Astronomy \\ The Johns Hopkins University
%\\ Baltimore, MD \\  }
% 
% 
%\affiliation{Department of Physics \\ Houghton College \\ Houghton, NY \\ }
%
% 
% \affiliation{Center for Theoretical Physics \\ Massachusetts Institute of Technology \\
% Cambridge, MA\\}
%
% 
%\affiliation{Theoretical Physics Group \\ Fermi National Accelerator Laboratory \\  Batavia, IL}
%

\date{\today}% It is always \today, today,

%\begin{abstract}

\abstract{
We present a simple UV completion of Atomic Dark Matter (aDM) in which 
 heavy right-handed neutrinos decay to induce both dark and lepton
number densities.  This model addresses several outstanding cosmological problems: the matter/anti-matter asymmetry, the dark matter abundance, the number of light degrees of freedom in the early universe, and the smoothing of small-scale structure.  Additionally, this realization of aDM may reconcile the CoGeNT excess with recently published null results and predicts a signal in the CRESST Oxygen band. We also find that, due to unscreened long-range interactions, the residual \emph{un}recombined dark ions settle into a diffuse isothermal halo.
}
 
 %\end{abstract}

%\pacs{}% PACS, the Physics and Astronomy
\maketitle

\section{\label{sec:intro} Introduction}
\noindent While indirect cosmological observations provide abundant evidence for the existence of dark matter (DM) \cite{Bertone:2004pz,Komatsu:2008hk,Abazajian:2004tn}, terrestrial evidence of 
its particle nature has been elusive.  The identity of DM stands alongside several important open questions at the intersection of cosmology and particle physics including the missing anti-matter, the number of light degrees of freedom in the CMB \cite{Mangano:2006ur}, and the observed absence of small-scale structure \cite{1996ApJ462563N,Gilmore:2007fy}.
\par\bigskip\noindent
Recently, the \cog direct detection experiment \cite{Aalseth:2010vx} reported an excess of events in their low-recoil bins. If this excess is interpreted as evidence of a DM particle, the natural scale for its mass
is  $\sim\mathcal{O}(10\,\GeV)$.  Since this energy scale does not easily fit the so-called WIMP paradigm, the dark sector must generically be expanded to generate the right cosmological abundance.  
This can be accomplished with new light states as in \cite{Finkbeiner:2007kk, Pospelov:2007mp, Zurek:2008qg} or by relating the DM abundance to the SM baryon asymmetry as in \cite{Kaplan:1991ah,Nussinov:1985xr,Barr:1991qn,Barr:1990ca,Gudnason:2006ug,Dodelson:1991iv,Kuzmin:1996he,Fujii:2002aj,Kitano:2004sv,Farrar:2005zd,Kitano:2008tk,Kaplan:2009ag,Kribs:2009fy,Falkowski:2011xh,Luty:1992un}.  See \cite{Buckley:2011kk} for a thorough treatment of the constraints on such models. 

\par\bigskip\noindent
 Generic models of light DM are highly constrained by the null results of CDMS \cite{Ahmed:2010wy} and XENON10 \cite{Angle:2007uj,Angle:2009xb,Angle:2011th} experiments.  
 The CDMS collaboration has recently reanalyzed the CDMS II Germanium data with the detection threshold lowered to 2 keV.  This analysis excludes both the DAMA  \cite{Bernabei:2010ke} and CoGeNT preferred regions for WIMP DM. 
  XENON10 also claims to rule out the WIMP interpretation of DAMA and CoGeNT, though there is controversy over XENON's scintillation efficiency $\mathcal{L}_{\textrm{eff}}$ at low
   energies \cite{Collar:2010gg,Collaboration:2010er,Collar:2010gd}.  
  A theory which can explain the positive signals while evading all the constraints may require some or all of the following epicycles: additional dark
   species \cite{Essig:2010ye}, momentum dependent DM/SM interactions \cite{Chang:2009yt} or non-standard couplings to nucleons \cite{Feng:2011vu}  (see Ref.\,\cite{Chang:2010yk} for a thorough study).

\par\bigskip\noindent
In this note, we suggest that atomic dark matter (aDM) may answer a number of important, open questions in cosmology.  We find that aDM can generate the right DM abundance and baryon asymmetry, contains additional relativistic degrees of freedom and is capable of smoothing structure on much larger scales than conventional CDM candidates \cite{Kaplan:2009de}.    Furthermore, aDM may reconcile \cog with constraints from null experiments. We also
 find that the regions of aDM parameter space favored by CoGeNT are consistent with preliminary signals at CRESST \cite{CRESSTunpub}.  
Finally, we note that that the existence of both dark ions and atoms within aDM gives rise to a unique halo structure.  

\par\bigskip\noindent
Section \ref{sec:admrev} gives a brief overview of aDM; section \ref{subsec:lepto} extends the simple 
 framework to explain both the dark matter abundance and the SM baryon asymmetry via the mechanism recently proposed in \cite{Falkowski:2011xh}; section \ref{subsec:irspectrum}  describes and justifies the pattern of spontaneous symmetry breaking in the dark sector; section \ref{subsec:recomb} describes the recombination of multiple species of dark atoms; section \ref{sec:exp} reviews relevant direct detection signals, limits and constraints on the aDM parameter space; subsection \ref{sec:halos} includes a discussion of the novel aDM halo structure;  finally, section \ref{sec:disc} summarizes our results and outlines future directions.

\section{\label{sec:admrev} Review of aDM}
\noindent Atomic dark matter consists of four Weyl fermions - \textbf{E}, \textbf{E$^c$}, \textbf{P} and \textbf{P$^c$} - charged under two $U(1)$'s.  The first, $U(1)_D$, has vector couplings and is unbroken.  The second, $U(1)_X$, has axial-vector couplings and is spontaneously broken by the vev of $ \mathcal{X}$ which is also responsible for the masses of \textbf{E} and \textbf{P}.

\begin{table}[h]
\centering
\begin{tabular}{|c|c|c|}
	\hline
		&	$U(1)_D$		&	$U(1)_X$		\\
		\hline
	\EE	&	\!\!\!$-1$			&	\!\!\!$-1$			\\
	\EEc	&	$1$			&	\!\!\!$-1$			\\
	\PP	&	$1$			&	$1$			\\
	\PPc	&	\!\!\!$-1$			&	$1$			\\
	$ \mathcal{X}$&	$0$			&	$2$		\\
	\hline
	
\end{tabular}
\caption{Field content and $U(1)$ charges for aDM.}
\label{tab:charges}
\end{table}%

\noindent The axial gauge boson is kinetically mixed with SM $U(1)_Y$ through a coupling of the form \cite{Holdom:1985ag}
\be
\label{eq:kineticmix}
\mathcal{L}_{\textrm{mix}} = \frac{\epsilon}{2} B_{\mu\nu}X^{\mu\nu}.
\ee
This operator arises from integrating out a heavy fermion with vector couplings to both $U(1)$'s so $\epsilon$  is given by:
\be
\label{eq:mixing}
\epsilon(\mu) = \frac{g_Y g_X}{16\pi^2}\ln{\left(\frac{M_{\textrm{heavy}}}{\mu}\right)} ~~,
\ee
where experimental constraints allow $\epsilon^2 \lesssim 10^{-5}$ for $M_X \gtrsim 400 \MeV$ \cite{Pospelov:2008zw,Bjorken:2009mm,Redondo:2010dp}. Note
that the existence of a \uone gauge boson with this mass and coupling can ameliorate the discrepancy between the standard model prediction and the 
measured value of the muon g-2 \cite{Pospelov:2008zw}. The field content and interactions above are capable of producing a successful cosmology and unique direct detection spectrum.

\subsection{\label{sec:admcosmo} Cosmology}
  The possibility of \uone~charged DM with long-range interactions, has been explored in a number of works \cite{Kaplan:2009de,Feng:2009mn, Ackerman:2008gi, Kaloper:2009nc, Dai:2009hx}.  In the case where DM exists in ionic form, halo morphology and bullet-cluster observations \cite{Markevitch:2003at,Randall:2007ph} place tight constraints on the $(\alpha_D, m_{DM})$ parameter space.  Long-range interactions push the DM from a virial configuration toward kinetic equilibrium and can make the scattering rate in the bullet cluster too high.  The aDM scenario avoids these problems by assembling the dark ions into atomic bound states which are \textit{net neutral} under the \uone~with a smaller fraction $X_E$ existing in ionic form.  The ionic fraction is defined as:
\be
X_E \equiv \frac{n_E}{n_E + n_H},
\label{ionfrac}
\ee
and it is most sensitive to the value of $\alpha_D$, tending to decrease as the coupling increases.  Similarly, $X_E$ also tends to decrease as $m_E$ increases with $m_{P}$ held fixed.  The dependence on $m_P$ is much weaker than the other two parameters.  See Figure \ref{fig:recomb} for the light atoms considered in this work and Figure 1 in Ref. \cite{Kaplan:2009de} for a more general treatment.
\par\bigskip 
In this framework, the cosmological abundance of DM is dependent upon the existence of an asymmetry between $(E,P)$ and $(E,P)^c$ and we return to the question of generating this asymmetry in Section \ref{sec:asym}.

\subsection{\label{sec:admsignal} Direct Detection}
  The leading interaction between aDM and the SM is through the $X - \gamma$ mixing in Eq.\,(\ref{eq:kineticmix}).  The static potential between a SM particle with charge $Q_{\textrm{EM}}$ and a DM ion with charge $Q_X$ goes like
\be
\label{eq:staticpotential}
V(\vec{S}_{\textrm{DM}},\vec{r}\,) \sim \left(\epsilon \,Q_X Q_{\textrm{EM}}\right)\,\left(\vec{S}_{\textrm{DM}} \cdot \vec{r}\right)\,\frac{e^{-M_X r}}{r^2},
\ee
with the dependence on the DM spin-operator arising from the axial-vector couplings of $\uone_X$, cf.\,Ref.\,\cite{Alves:2009nf}.  As in SM hydrogen, the aDM ground state is the $n = 1$ state with anti-aligned spins and the $S = 1$ triplet states have a slightly higher energy so there is a hyperfine splitting.  At leading order, the interaction in Eq.\,(\ref{eq:staticpotential}) forces dark atoms to scatter \textit{inelastically} from SM nuclei by excitation into the hyperfine state.  The ratio of the hyperfine splitting $E_{\textrm{hf}}$ to the ground state binding energy $B$ scales as
\be
\frac{E_{\textrm{hf}}}{B} \propto \alpha_D^2 \frac{m_E}{m_P},
\ee
so that $E_{\textrm{hf}}$ can easily be $\mathcal{O}(\keV)$ for atomic masses $\mathcal{O}(10 \,\GeV)$.  
%Furthermore, the linear dependence on $\vec{r}$ in Eq.\,(\ref{eq:staticpotential}) makes the  scattering rate linear in the nuclear recoil energy and tends to 
%push .  
This implies that dark ions, which are free spins, will scatter \textit{elastically} such that the ionic recoil spectrum vanishes for small recoil energies.  Thus, aDM realizes many of the mechanisms \cite{Chang:2010yk} necessary for reconciling \cog with other null searches.  In Section \ref{sec:exp} we show that aDM 
can explain the positive signals reported by both \cog and CRESST while evading bounds set by XENON and CDMS.

%%%%%%%%%%%%%%%%%%%%%%%%%%%%%%%%%%%%%%%%%%%%%%%%%%%%%%%%%%%%%%%%%%%%%%%%%%%%%%%%%%%%%%%

%											Model Introduction and connection to Leptogenesis
			
%%%%%%%%%%%%%%%%%%%%%%%%%%%%%%%%%%%%%%%%%%%%%%%%%%%%%%%%%%%%%%%%%%%%%%%%%%%%%%%%%%%%%%%

\section{\label{sec:asym}  Asymmetric Atomic Dark Matter}

In this section we propose an ultraviolet completion to the above model. It both dynamically explain the generation of the dark matter 
abundance (by linking it to the baryon asymmetry), and relieves the issue of a Landau pole for the \uone\,  dark gauge field below the 
Planck scale. 

\subsection{\label{subsec:model} The Model}
%Atom is from the greek word 'atomos'...is this way too nerdy?%
  We propose a nonabelian dark sector with $SU(2)_{D} \times U(1)_{X}$ gauge symmetry, where the labels $D$ and $A$ refer to ``dark'' and ``axial,'' respectively.  By embedding $U(1)_D$ into a non-Abelian group we avoid a Landau pole below the Planck scale.  The matter Lagrangian contains
\be
\hspace{-0cm}
\!\!\!\!\!\mathcal L \! &\supset& -\frac{1}{2}  M_{\!n}^{i} n_{i}^{2} + y^{ij} n_{i} \ell_{j} h +  \lambda_{\textrm{\textbf{e}}}^{i} n_{i} E \varphie + \lambda^{i}_{\textrm{\textbf{p}}} n_{i} P \varphip + y_{e} \, \mathcal{X} E E^{c}
 +  y_{p}\,  \mathcal{X}^{\dagger} P P^{c} + {\rm H.c.}~~,
 \label{eq:Lagrangian}
\ee 
  where $\ell_{j}, h$ are the Standard Model lepton and Higgs doublets; the $n_{i}$ (for $i = 1,2$) are sterile neutrinos with GUT scale Majorana masses $M_{i}$; the 
$\varphip, \varphie,$ and $  \mathcal{X}$ are scalar fields. All gauge representations and quantum numbers are given in Table \ref{tab:charges2}. 

\begin{table}[h]
\centering
\begin{tabular}{|c|c|c|c|}
	\hline
		&	$SU(2)_D$		&	$U(1)_X$		&		$\mathcal{Z}_{2}$	\\
		\hline
	\EE	&	$\overline\Box$			&	\!\!\!\!$-1$			&		  \!\!\! $-1$	\\
	\EEc	&	$\Box$			&	\!\!\!\!$-2$			&						 \!\!\!   $-1$		\\
	$\varphie$ &    $\Box$                        &       $1$                     &                                           \!\!\!$-1$                      \\
	\PP	&	$\Box$			&	$1$			&						    $1$		\\
	\PPc	&	$\overline\Box$			&	$2$			&						   $1$		\\
	$\varphip$ &    $\overline\Box$                        &  \!\!\!\!\!$-1$                     &                                           $1$                      \\
	$ \mathcal{X}$&    $ \begin{array}{cc} \Box&  \!\!\! \Box  \vspace{-0.22cm}\\   \Box & \end{array} $& $ 3$			&						$1$			\\  
			\hline
\end{tabular}
\vspace{0.3cm}
\caption{Field content and gauge representations for Asymmetric aDM. The \uone$_{X}$ charge assignments forbid $nE^{c} \varphie$ and $nP^{c} \varphip$ terms which 
would wash out the dark matter asymmetry. The discrete $\mathcal Z_{2}$ parity prevents atomic annihilation in the low energy effective theory.  
Mixing between \uone$_{Y}$ and \uone$_{D}$ is naturally tiny due to the  $SU(2)_{D} $ embedding. 
} 
\label{tab:charges2}
\end{table}%

\par \bigskip
For at least two species of sterile neutrinos, the parameters $y^{ij}$ and $\lambda_{\bf e,\,p}^{i}$ contain irreducible complex-phases and give rise to CP violation. 
Out of equilibrium $n$ decays generate both the Standard Model lepton asymmetry and  the asymmetric dark matter abundance. 
While lepton number is explicitly violated by  
neutrino Majorana masses, it remains a good 
accidental symmetry  in the visible sector above the electroweak scale. In the dark sector, we impose a  ${\cal Z}_{2}$ symmetry  to dangerous $EP$ mass terms which allow 
$E P$ annihilation into dark radiation, see Table \ref{tab:charges2}.  Notice that Eq.\,(\ref{eq:Lagrangian}) does not allow explicit mass terms for the fermions $E$ and $P$; 
however, dark-sector symmetry breaking
via the VEV $\left<  \mathcal{X} \right> \equiv v_{ \mathcal{X}}$ induces these fermion masses through the $ \mathcal{X} E E^{c}$ and $  \mathcal{X}^{\dagger} PP^{c}$ yukawa terms, as we will see in Section \ref{subsec:irspectrum}.  

\subsection{\label{subsec:lepto} Connecting Atomogenesis to Leptogenesis}
Following \cite{Falkowski:2011xh} and \cite{Luty:1992un}, we track the evolution of these asymmetries with the parameters
\be
\label{eq:DefineAsymm}
\epsilon_\ell &=& \frac{\Gamma \left( n_1 \rightarrow l h \right)-\Gamma \left( n_1 \rightarrow \bar l h^\dagger \right)}{\Gamma_{n_1}} \\
\epsilon_{E} &=& \frac{\Gamma \left( n_1 \rightarrow E \varphie   \right)-\Gamma \left( n_1 \rightarrow \bar E \varphi_{\bf e}^\dagger \right)}{\Gamma_{n_1}}\\ 
\epsilon_{P} &=& \frac{\Gamma \left( n_1 \rightarrow P \varphip   \right)-\Gamma \left( n_1 \rightarrow \bar P \varphip^\dagger \right)}{\Gamma_{n_1}} ~~,
\ee
Since the IR phenomenology will require $E$ and $P$ to be stable with comparable masses, we will simplify our discussion by considering only
the asymmetry in $E$ without loss of generality.
%For generic phases in the $\lambda$ and $y$ couplings, the decay-asymmetries ratio satisfies
%\be
%\frac{\epsilon_{\ell_{j}}}{\epsilon_{E}} \simeq \frac{ y^{1j}_{\bf e} | y^{2j}_{\bf e}|      }{\lambda^{1}_{\bf e} |\lambda^{2}_{\bf e}|}  .
%\ee
%Following \cite{luty,falkowski} in the hierarchical limit where $M_{N_{1}} \ll M_{N_{2}} $ and defining the ``yield''  $Y_{i}\equiv n_i/s$
for each number density of interest,
the yields $Y_{i} \equiv n_{i}/s$ satisfy the Boltzmann equations
\begin{eqnarray}
\label{eq:BE1}
\hspace{-0.5 in}\frac{s H_1}{z} Y_{n_1}' &=& -\gamma_D \left( \frac{Y_{n_1}}{Y_{N_{1}}^{\rm \,eq}} -1 \right)+\, (2 \leftrightarrow 2) \,,
\\
\label{eq:BE2}
\frac{s H_1}{z} Y_{\Delta E}' &=& \gamma_D \left[ \epsilon_{E}\!\left( \frac{Y_{n_1}}{\,Y_{n_1}^{\rm\, eq}} \!-\!1 \right) - \frac{\!Y_{\Delta E}}{\>2 Y_{E}^{\rm \,eq}} \,
\mathcal{B}_{E} \right] + (2 \leftrightarrow 2 ~\textrm{washout + transfer})
\\
\label{eq:BE3}
\frac{s H_1}{z} Y_{\Delta \ell}' &=& \gamma_D \left[ \epsilon_\ell \left( \frac{Y_{n_1}}{Y_{n_1}^{\rm \, eq}} -1 \right) - 
\frac{\!Y_{\Delta \ell}}{\>2 Y_\ell^{\rm \,eq}} \, \mathcal{B}_\ell\right] + (2 \leftrightarrow 2 ~ \textrm{washout + transfer} ) ~,
\end{eqnarray}
where $'$ denotes differentiation with respect to $z \equiv M_{n_1}/T$, $\Delta_{(\ell, E)}$ track the particle-antiparticle asymmetries in the two
sectors,  $H_1$ is the Hubble parameter at $T = M_{n_1}$, $s$ is the total entropy density, $Y_{i}^{\rm eq} $ are the equilibrium yields,
%  $Y_x^{\rm eq} = g_x z^2 K_2(z)/8 g_*$,
 $\mathcal{B}$ denote the branching fractions of $n_1$ into the corresponding channel
and finally, $\gamma_D$ is  the thermally averaged $n_1$ decay density
\be
\label{eq:GammaD}
\gamma_D = \frac{m_{n_1}^3  K_1(z)} {\pi^2 z}  \Gamma_{n_1}~~,
\ee
which we have written in terms of the first modified Bessel function $K_{1}$. 
\par\bigskip
In order to generate the observed scale of neutrino masses $\mathcal O(10^{-2} \,\eV)$ via the ``See-Saw'' mechanism 
and the correct abundance of 
$\mathcal O(10 \,\GeV)$ dark matter, we 
must work in the so-called ``strong-strong'' washout regime where both SM and 
dark sector partial-widths satisfy   $\mathcal{B}_{\ell,E} \Gamma_{n_1}^2  \gg M_{n_1} H(M_{n_1})$.
In this scenario the 
neutrinos remain coupled to the cosmological fluid until 
the  $2\leftrightarrow 2$ scattering terms (e.g. $n_{1} n_{1} \leftrightarrow \ell \ell $) trigger the departure 
from equilibrium 
after the neutrino number density becomes nonrelativistic. This allows $Y_{n_{1}}$ to drift from $Y^{\mathrm eq}_{n_{1}}$ 
and leave behind asymptotic particle/antiparticle asymmetries $Y^{\infty}_{\Delta (\ell, E)}$ in the $z \to \infty$ limit. 

\begin{figure}[t]
%\vspace*{0.2cm}
\begin{center}
\unitlength=0.7 pt
\hspace*{-1cm}
\SetScale{0.7}
\SetWidth{0.9}      % line    size control
\normalsize    %  letter  size control
{} \allowbreak
%  diagram # 1
\begin{picture}(100,90)(130,0)
\DashArrowLine(15,30)(70,30){4}
\ArrowLine(110,50)(70,30)
\Line(103,53)(117,48)
\Line(114,58)(107,43)
\ArrowLine(109,50)(142,67)
\ArrowLine(135,-2)(70,30)
\Text(22,16)[c]{\small $ \varphie$}
\Text(78,54)[c]{\small $ n $}
\Text(160,71)[c]{\small $ \nu ,  \ell$}
\Text(151,-4)[c]{\small $\overline E$}
\put(211,0){
\DashArrowLine(15,30)(70,30){4}
\ArrowLine(110,50)(70,30)
\Line(103,53)(117,48)
\Line(114,58)(107,43)
\ArrowLine(109,50)(142,67)
\ArrowLine(135,-2)(70,30)
\Text(22,16)[c]{\small $ \varphip$}
\Text(78,54)[c]{\small $ n $}
\Text(160,71)[c]{\small $ \nu ,  \ell$}
\Text(151,-4)[c]{\small $\overline P$}
}
\end{picture}
\end{center}
\vspace*{-0.2cm}
\caption{Diagrams contributing to scalar doublet decay through 
neutrino mass insertions. After the scalars become matter-antimatter symmetric 
 through $\varphie$ and $\varphip$ number
violating interactions, these decays give no {\it net} lepton number
violation and the decay products annihilate into dark/visible radiation. }
\label{fig:decay}
\end{figure}
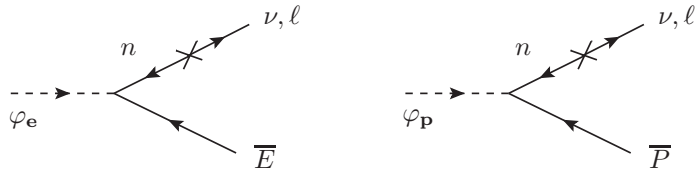

The Lagrangian in Eq.\,(\ref{eq:Lagrangian}) only displays terms that exhibit a global symmetry under which $E (P)$ and $\varphie (\varphip)$ 
carry opposite charge. After electroweak symmetry breaking, the scalars $\varphi_{\bf e, p}$ can decay to $(\ell \bar E)$ and $(\ell \bar P)$ final states (Figure \ref{fig:decay}).  Since the scalars acquire particle-antiparticle excesses equivalent to their fermionic counterparts,  their decays naively erase the asymptotic fermion asymmetry $Y^{\infty}_{\Delta E}$.  However, the scalar potential for these fields allows terms that  violate $\varphie$ and $\varphip$ number by two units
\be
V(\varphip, \varphie) \supset  \kappa\, (\varphip \varphie)^2 + {\rm h.c.} ~~,
\label{eq:phiconvert}
\ee
and thereby initiate interconversion $\varphi_{\bf e,p}\leftrightarrow \varphi_{\bf e,p}^{\dagger}.$  
When the dark asymmetry acquires its asymptotic value at  $T_{\mathrm{asym}} \gg M_{\varphi}, $
the scalars are still relativistic and the interactions in Eq.\,(\ref{eq:phiconvert}) equilibrate with the thermal bath
to washout the scalar asymmetry before they decay out of equilibrium\footnote{Technically this requirement is 
too strong; the decay need not necessarily be out of equilibrium, but this is generically the case 
since the only allowed process (Figure \ref{fig:decay}\,) is suppressed by powers 
of $ v/M_{n_{1}}$ and becomes relevant only after interconversion
has frozen out.  } at late times.
 Since there is no comparable interaction for $E$ or $P$, the resulting 
dark sector will only contain stable asymmetric fermions.

 %%%%%%%%%%%%%%%%%%%%%%%%%%%%
%      Decay Modes  Diagrams  --<  
 %%%%%%%%%%%%%%%%%%%%%%%%%%%%

%%%%%%
 
\par\bigskip As with standard Leptogenesis, electroweak sphalerons generate the observed baryon number from the lepton asymmetry at high temperatures.
If the Yukawa couplings $|\lambda|$ and  $|y|$ are identical in magnitude and phase,  
  then both sectors acquire the same  particle-antiparticle asymmetries. The ratio $\Omega_{DM}/\Omega_{B}$
 will therefore have the observed value of $\simeq 6$ if the average mass in the dark sector is $\mathcal O(10 \, \GeV)$; we will assume this to be the case throughout the remainder of this paper. 
 
 \par\bigskip Finally, we note that in the limit where we ignore all interactions not included in Eq.\,(\ref{eq:Lagrangian}), we can define
 \be
 \sigma &\equiv&  \begin{pmatrix} \sigma_+\\ \sigma_- \end{pmatrix} \equiv \sqrt2 \begin{pmatrix} \tilde{\lambda}_{\textrm{\textbf{e}}} E + \tilde{\lambda}_{\textrm{\textbf{p}}} P\\ \tilde{\lambda}_{\textrm{\textbf{e}}} E - \tilde{\lambda}_{\textrm{\textbf{p}}} P \end{pmatrix}\nonumber\\
 \varphi &\equiv& \begin{pmatrix} \varphi_+\\ \varphi_- \end{pmatrix} \equiv \sqrt2 \begin{pmatrix} \varphie + \varphip\\ \varphie - \varphip \end{pmatrix},
 \ee
 so that Eq.\,(\ref{eq:Lagrangian}) contains
 \be
 \lambda^i n_i\,\sigma \cdot \phi,
 \ee
 where $\tilde{\lambda}_{\textrm{\textbf{e,p}}} \equiv \frac{\lambda^i_{\textrm{\textbf{e,p}}}}{\lambda^i}$.  Thus, we see explicitly that our UV theory is physically identical to that in \cite{Falkowski:2011xh}, which finds robust
parameter space for thermal ``See-Saw'' Leptogenesis with dark matter mass $m_{\chi} \simeq 10 \,\GeV$.  
Since the model's IR features (e.g.\,direct detection, structure formation) are not sensitive to the parameters in
 the UV Lagrangian, in the rest of the paper we take the asymmetry for granted. Furthermore, we will assume that 
 the couplings $\lambda_{\textrm{\textbf{e}}}^{i}$ and $\lambda_{\textrm{\textbf{p}}}^{i}$ in Eq. (\ref{eq:Lagrangian}) are 
 such that the resulting asymmetries give equal numbers of $E$ and $P$ states at late times.

\subsection{\label{subsec:irspectrum} Symmetry breaking and IR mass spectrum }
 
The scalar potential for the adjoint $ \mathcal{X}$ contains
\be
V( \mathcal{X}) \supset \eta \,( \mathcal{X}^{a \dagger}  \mathcal{X}^{a})^{2} + \eta^{\prime} \, \mathcal{X}^{a \dagger}  \mathcal{X}^{b}  \mathcal{X}^{a \dagger} \mathcal{X}^{b} + M_{ \mathcal{X}}^{2}  \mathcal{X}^{a \dagger}  \mathcal{X}^{a}  ~,~
\ee
 
  where $a$ and $b$ are $SU(2)_{D}$ adjoint indices. While couplings to the other scalars are also allowed, we demand that 
$\left<\varphie\right> = \left<\varphip \right> = 0$,  so operators with these fields do not contribute to the minimization conditions. 
We have also omitted the allowed SM Higgs coupling $H^{\dagger }H \mathcal {X}^{a \dagger}  \mathcal {X}^{a}$ and absorbed its
  vev into $M_{\mathcal X}$ for simplicity.
\par\medskip
% 
%\be
%\frac{\partial V}{\,\,\partial  \mathcal{X}^{b\, \dagger}} =  2\eta ( \mathcal{X}^{a \dagger}  \mathcal{X}^{a})  \mathcal{X}^{b}  + 2 \eta^{\prime} ( \mathcal{X}^{a}  \mathcal{X}^{a})  \mathcal{X}^{b \,\dagger} + 
%M^{2}_{\mathcal X} \,  \mathcal{X}^{b}  =  0  ~,
% \ee
 
   For $ M_{\mathcal X}^{2} <0$,
 the adjoint scalar acquires a VEV which we can rotate into the $T_{3}$ direction without loss of generality
 \be
 \left<  \mathcal{X}^{3}\right> =  \left<  \mathcal{X}^{3 \dagger}\right>  \equiv v_{\mathcal{X}}= \sqrt{\frac{ M_{\mathcal X}^{2}}{2(\eta + \eta^{\prime})}  } ~.
 \ee
 Since $ \mathcal{X}$ is an $SU(2)_{D}$ doublet with $U(1)_{X}$ charge, 
 this implies a symmetry breaking pattern where the axial group is broken completely $SU(2)_{D}\times U(1)_{X} \rightarrow U(1)_{D}$,
  while the residual unbroken $U(1)_{D}$ is just the $T_{3}$ component of $SU(2)_{D}$. Henceforth, we will refer to this massless gauge field 
 as the ``dark photon.''  

\par\bigskip 
After symmetry breaking, the fermionic doublets $E, P$ acquire masses $m_{E, P} \equiv y_{\bf e, p} v_{\mathcal X}$ and residual $U(1)$ charges are determined by their $SU(2)_{D}$ isospin.
\be 
E \equiv
 \left( \begin{array}{ccc}
{\tilde{e}}  \\
{e} \end{array} \right)   \quad\quad\quad, \quad \quad \quad
P \equiv
 \left( \begin{array}{ccc}
{p}  \\
{\tilde{p}} \end{array} \right)
\ee

As noted previously, gauge charges allow an $EP$ mixing mass, which would allow atomic states to annihilate, hence we demand a $\mathcal{Z}_{2}$ symmetry to forbid this mixing and stabilize our dark matter candidate.

\subsection{\label{subsec:recomb} Recombination of Multiple Atomic Species }
For sufficiently large dark couplings (e.g. $\alpha_{D}\sim 0.1$),  aDM gives robust parameter space for early-universe recombination. 
The original scenario, however, assumes the minimal field content giving rise to only one species of atom: a Hydrogen-like bound state with hierarchical constituents (e.g. $m_p \sim 100\,m_e$).
In the $SU(2)_{D}\times$ \uone$_{X}$ model, the field content allows four distinct  atomic bound states.   After $ \mathcal{X}$ acquires a VEV,  dark ``electrons''  $E$  and dark ``protons''  $P$ generically receive different masses. Since 
both doublets have charge $\pm 1$ components  $(\tilde{e}, e)$ and $(p, \tilde{p})$ under the unbroken $U(1)_{D}$  symmetry, predicting the cosmological atomic abundance requires following the evolution of 8 correlated species: $\tilde{e},\,e,\,p,\,\tilde{p},\,H_{ep},\,H_{\tilde{e}\tilde{p}},\,H_{e\tilde{e}}\;\textrm{and}\;H_{p\tilde{p}}$.  The residual $SU(2)_D$ \emph{global} symmetry guarantees that tilded and un-tilded fields evolve in the same way, which reduces the number of independent species to five.  Finally, we can reduce the number of independent equations to four if we demand that the co-moving DM number density is constant, where

\be
\label{eq:ndm}
n_{DM} = 2n_e+ 2n_p+4N_{ep}+2N_{e\tilde{e}}+2N_{p\tilde{p}}.
\ee  
If we define the following fractional yields
\be
X_e n_{DM} &=& 2 n_e \nonumber\\
X_p n_{DM} &=& 2 n_p \nonumber\\
Y_{ep} n_{DM} &=& 2 N_{ep} \nonumber\\
Y_{p\tilde{p}}\,n_{DM} &=& 2 N_{p\tilde{p}}\nonumber \\
Y_{e\tilde{e}}\,n_{DM} &=& 2 N_{e\tilde{e}},
\ee
then Eq.\,(\ref{eq:ndm}) becomes
\be
1=X_e+ X_p + 2Y_{ep} + Y_{e\tilde{e}} + Y_{p\tilde{p}}.
\ee
Without loss of generality we set $Y_{e\tilde{e}}=1-X_e- X_p - 2Y_{ep}-Y_{p\tilde{p}}$\;and take the independent Boltzmann equations to be\footnote{In the rest of this discussion we assume that $CP$-violation is negligible, i.e.\,the matrix elements in these Boltzmann equations are $T$-invariant.}
\be
\frac{d X_e}{dt} &=& \frac{2}{n_{DM}}(C_{ep}+C_{e\tilde{e}}) \nonumber\\
\frac{d X_p}{dt} &= &\frac{2}{n_{DM}}(C_{ep}+C_{p\tilde{p}}) \nonumber\\
\frac{d Y_{p\tilde{p}}}{dt} &=& -\frac{2}{n_{DM}}(C_{p\tilde{p}}) \nonumber\\
\frac{d Y_{ep}}{dt} &=& -\frac{2}{n_{DM}}(C_{ep}) \\
\ee
The collision operator $C_{ij}$ for the recombination of ions $i$ and $j$ into bound state $H_{ij}$ can be written as
\be
C_{ij} = \langle\sigma\rangle_{ij\rightarrow H_{ij}\gamma} \left(N_{ij}\frac{n_i^{\textrm{eq}}n_j^{\textrm{eq}}}{N_{ij}^{\textrm{eq}}}-n_in_j\right);
\ee
the superscript ``eq'' refers to equilibrium number density and the full expression for the thermally averaged recombination cross-section can be found in our earlier paper \cite{Kaplan:2009de} and references therein.

\begin{figure}[t]
\centering

\subfigure[ \,Dashed lines are constant $E_{\textrm{hf}}$ in keV.]{\label{fig:recombhyper}\includegraphics[width=0.4\textwidth]{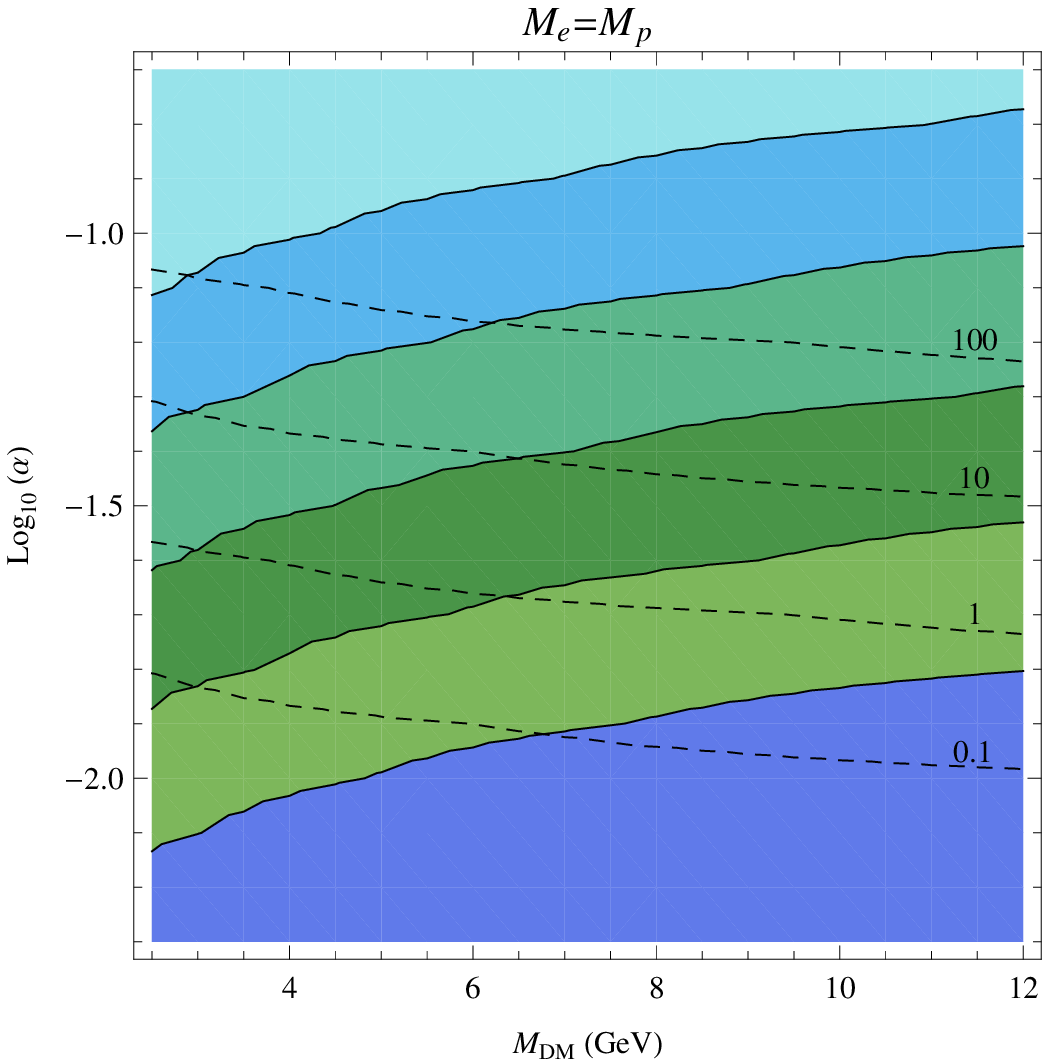}}
\subfigure[ \,Dashed lines are constant $\sigma_{\textrm{self}}/M_{\textrm{DM}}$.]{\label{fig:recombscatter}\includegraphics[width=0.4\textwidth]{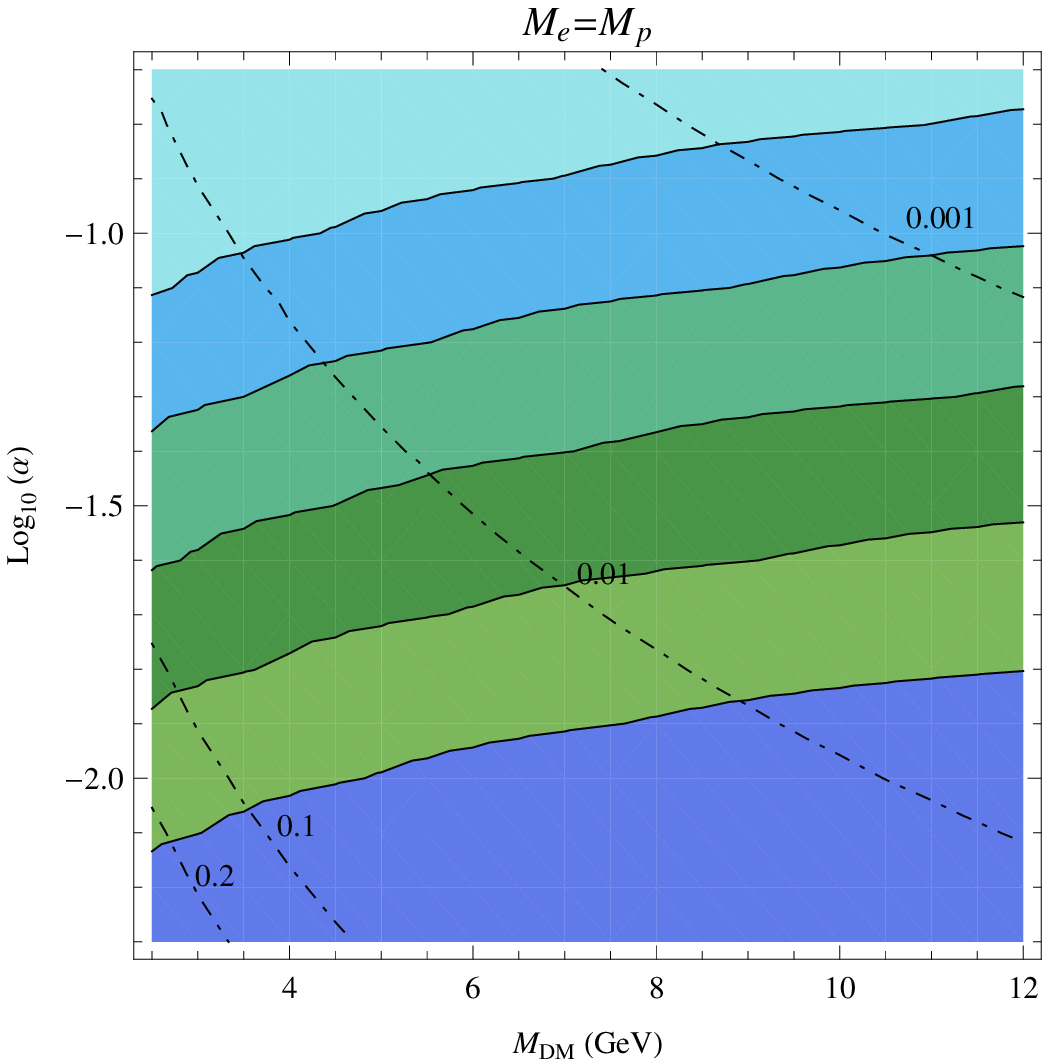}}

\caption{Solid lines show the total residual ionization. In both plots the values of the residual ionization $X_e$ are, from top to bottom: $> 0.1$, $10^{-2}-10^{-1}$, $10^{-3}-10^{-2}$, 
$10^{-4}-10^{-3}$, $10^{-5}-10^{-4}$ and $10^{-6}-10^{-5}$. Dashed lines on the left plot indicate the hyperfine splitting in keV, while dashed lines on the right plot indicate constant values of the ratio of the self-scattering cross section to the dark matter mass
in ${\rm cm}^{2}/$ GeV. In both cases, the horizontal axis is the total mass of the dark atom.
}

\label{fig:recomb}
\end{figure}

\par\bigskip
While the total dark matter number density depends on the abundances of all species, the ``chargitronium'' states $e\tilde{e}\;\textrm{and}\;p\tilde{p}$ do not interact with ordinary matter at leading order; see Section \ref{sec:dirdet} for a detailed discussion.  In Figure \ref{fig:recombhyper} we plot the fractional cosmological abundance of atomic states $2Y_{ep}$ as a function of $\alpha_D$ and the atomic  mass $m_{DM}$, including contours of constant hyperfine splitting.  
Observations of the bullet cluster and constraints from DM halo morphology 
(see Section 2.1 in \cite{Kaplan:2009de}) demand that $\sigma_{\textrm{self-scattering}}/M_{\textrm{DM}} \lesssim$ 1 cm$^2$/g.  The self-scattering cross sections for ion-atom and atom-atom interactions are large -- larger than the na\"{i}ve geometric value $4\pi a_0^2$ --  because the interaction potentials are long-range, mediated by the massless dark photon.  As noted in \cite{Kaplan:2009de} and references therein, over the relevant range of interaction velocities we consider, the actual cross-sections scale as $\sigma_{\textrm{self-scattering}} \sim 4\pi (\kappa \,a_0)^2$, where $3 \le \kappa \le 10$ sets the scattering length. In Figure \ref{fig:recombscatter} we plot the same parameter space with contours of constant $\sigma_{\textrm{self-scattering}}/M_{\textrm{DM}}$.  For the rest of the paper we will focus on the
regions of  parameter space where $2Y_{ep} \sim \mathcal{O}(1)$ and $X_e + X_p \leq 10\%$.

\subsection{\label{subsec:dof} Light Degrees of Freedom and the CMB}
The CMB is sensitive to the number of relativistic degrees of freedom in equilibrium with the photon gas, parameterized as the effective number 
of neutrinos, $N_{\nu}$
\be
\rho_{\rm rad} = \rho_{\gamma} + \rho_{\nu} + \rho_{\gamma_{\rm dark}} =  
\left[1 + \zeta \,\, \frac{7}{8} \left(\frac{4}{11}\right)^{4/3}\! N_{\nu} \right] \rho_{\gamma}~~,
\label{}
\ee
where $ \rho_{\gamma_{\rm dark}}$ is the radiation density due the dark photon and $\zeta \simeq 0.93$ is a parameter that 
corrects for neutrino/electron scattering and finite-temperature QED effects \cite{Ichikawa:2008pz}. 
The dark photon and ordinary photon are equilibrated by dark-electron visible-electron scattering through $X$-boson exchange,
which  becomes inefficient when the dark electrons become non-relativistic.  Their number density quickly becomes Boltzmann
 suppressed and the the two sectors decouple around the temperature $T_{\textrm{dec}} \approx m_{e}/20$, where $m_{e}$ is 
 the mass of the dark electron.  Once the dark/visible photon gasses decouple, they maintain relativistic number densities, so any temperature difference
 that arises between them is due entirely to the additional freeze-out of relativistic species, which 
 heats the visible radiation. 
\par\bigskip
The dark photon's contribution to $N_{\nu}$ in the CMB depends strongly on whether the sectors decouple 
before or after the QCD phase transition.  For dark electron masses at our scale of interest ($\sim$ 1 GeV),
 the sectors decouple around 50 MeV $\ll \Lambda_{QCD}$, so the visible sector only gets reheated
 by standard model electron, positron and neutrino freeze-out. Between decoupling and
last scattering, approximately 10 relativistic degrees of freedom freeze out in the visible sector, so the ratio 
of photon densities is 
\be
\frac{\rho_{\gamma_{\rm dark}}}{\rho_{\gamma}}  = \left(\frac{8}{43}\right)^{4/3} ~~,
 \ee
 which gives $N_{\nu} \simeq 3.4$  at last scattering in the presence of dark radiation. 

\section{\label{sec:exp} Direct Detection and Allowed Parameter Space}

\subsection{\label{sec:halos} Isothermal Ionic Halo }

In this section we consider the fate of dark ions that survive early-universe recombination.  For simplicity, we will assume single species of dark electrons $E$ and protons $P$. In the equal mass limit, $m_{E} = m_{P}$, this assumption introduces no loss of generality and the qualitative features of this argument do not change so long as the electron and proton masses are of the same order of magnitude. To model the cold DM, luminous disk, and bulge, we follow
the discussion in \cite{Catena:2009mf}, however our qualitative results are robust under perturbations of model input parameters
and persist when we consider different CDM haloes (e.g. NFW). 
\par\bigskip
In the allowed regions of aDM parameter space, atomic bound states are the dominant form of DM and both atom-atom and atom-ion scattering rates are suppressed.  As such, we can safely 
suppose that the CDM atoms in our galaxy settle into an Einasto\footnote{The qualitative results of this section 
do not change when we use the NFW profile \cite{1996ApJ462563N} to model the dominant atomic CDM halo.} profile \cite{1989A&A22389E} at late times
\be
\rho_{\rm atom} (r) = \rho_{\odot} \, { \exp}\left\{-\frac{2}{\alpha_e} \left[\left(\frac{r}{a_h}\right)^{\alpha_e} -
\left(\frac{r_{\odot}}{a_h}\right)^{\alpha_e} \right] \right\}
\ee
where $\rho_{\odot} = 0.3\, \GeV/ \rm{cm}^{3}$ is the local DM mass density, the Einasto index is $\alpha_{e} = 0.22$, and the length scale is  $a_{h} = 13 \,{\rm kpc}$.
We assume that the presence of dark ions does not significantly alter the CDM profile.  The luminous disk can be modeled as 
\be
\rho_d (r, z) =  \frac{\Sigma_d}{2 z_d} \,{ \exp} \left(-\frac{r}{r_d} \right) {\rm sech}^2 \left(\frac{z}{z_d}\right)
\ee
where $(r,z)$ are cylindrical coordinates,  $\Sigma = 1154 \,M_{\odot} / {\rm pc}^2$ is the surface density, 
 and $r_d = 2.54 $ kpc  ($z_{d} = 0.34$) is the radial (axial) scale factor. Finally, the luminous ``bulge" can 
 be modeled as a uniform sphere centered at the galactic origin. Since this 
lies well within the solar radius, our model will be insensitive to the bulge profile, so the total
bulge mass enclosed in radius $r$ is 
\be
M_{b}(r) = M_{b}\left(\frac{r}{r_{b}}\right)^3
\ee
where $M_b = 4.5 \times 10^9 M_{\odot}$ and $r_b =  1.54$ kpc.

\par\medskip

Although recombination leaves behind a global ionized fraction $X_{E}$ (see Eq.\,(\ref{ionfrac})),
 after galaxy formation, the dark-ion mass distribution inside the halo can deviate significantly from a standard profile.
To investigate this phenomenon, we assume a  conservative initial condition in which the ions are initially distributed in an Einsasto profile $\rho_{\rm ion}(t = 0; r) = X_{E} \,\rho_{\rm atom}(r)$, which
becomes distorted as they scatter.  
 While this approach does not take into account the initial ionic power spectrum, it sets an upper bound on the local  
 ionized fraction; ions encounter more friction during galactic infall and, therefore, comprise a smaller fraction of 
the total halo than our naive estimate ($\sim X_{E}$) would suggest.
\par\medskip

 Following the discussions in \,\cite{Feng:2009mn,Ackerman:2008gi} we consider the relaxation time
 $\tau$ for an ion to exchange an $\mathcal O(1)$ fraction of its kinetic energy. The classical scattering
 rate is 
 \begin{eqnarray}
\Gamma = n_{\rm ion}(r_{\odot}) \sigma\, v(r_{\odot}) = \frac{4 \alpha_{D}^{2} n_{\rm ion}(r_{\odot})     }{m_{\rm ion}^2 v^3}
\end{eqnarray}

where $n_{\rm ion}$ is the ion density and implicitly depends on $X_E$ and we have used the geometric cross section $\sigma \sim b^{2},$ where 
$b = 2 \alpha/ m_{\rm ion} v^{2}$ is the hard-scattering impact parameter . Comparing the relaxation time, $\tau = \Gamma^{-1}$ to the galactic period, we demand 
that a typical ion undergoes many hard scatters during the lifetime of the galaxy 
 \be
\frac{\tau}{T} \simeq \frac{G^{2} M(r_{\odot})^{2} m_{\rm ion}^{2}}{8\pi^{2} \alpha_{D}^{2} \,r_{\odot}^{3} \,n_{\rm ion}(r_{\odot})}\ll 50~~;
\label{eq:relax}
\ee
where $T=2\pi r_{\odot} /v$ is the galactic period and  $M(r)$ is the total (non-ionic) mass enclosed in radius $r$. In the parameter space we consider, this condition is 
trivially satisfied, and the ions reach kinetic equilibrium, settling into an independent isothermal halo. 

 \par\bigskip
The final equilibrium temperature of the ionic halo is set by a weighted average of the initial ionic speed distribution.  If we assume the ions are initially distributed virially,
 then by the virial and equipartition theorems, the temperature as a function of position is
\be
T(r) = \frac{G m_{\rm ion}}{3\, r} M(r)~~,
\ee
 where $M(r)$ is the total galactic mass enclosed in radius $r$. This gives an average temperature 
 \be
 \overline T =  \frac{1}{M_{\mathrm{G}}} \int  d^{3}r   \rho_{\mathrm{G}}(r) \,T(r)   ~~,
  \ee
  where  $\rho_{\mathrm{G}}$ and $M_{\mathrm{G}}$ are the galactic mass-density and total-mass respectively. The isothermal ion
  number density is, therefore 
  \be
  n_{\rm ion}(r) = {\cal C}\, e^{-\frac{ U(r)}{\overline T}} ~~,
  \ee
  where  $U(r)$ is the galactic gravitational potential\footnote{For $X_{E} \ll1, U(r)$ is approximately independent of the ionized fraction 
  so the result in Eq. (\ref{localfrac}) varies linearly with $X_{E}$.} and $\cal C$ is a normalization constant\footnote{Since ion velocities do not vary spatially at equilibrium, the kinetic term in the
  Boltzmann weight has been absorbed into the normalization. } set by the global ionized fraction $X_{E}$. \par\bigskip
  
  \begin{figure}
\begin{center}%
\hspace*{-2.5cm}
{\includegraphics{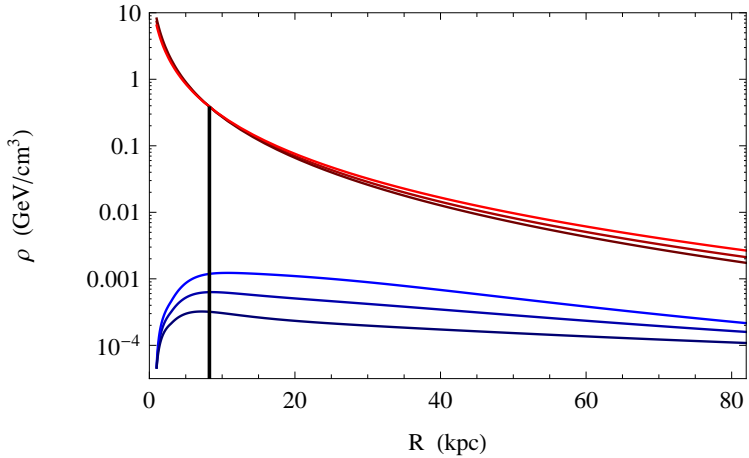}}

\caption{Plot of both atomic (red, higher) and ionized (blue, lower) mass densities as a function of distance from the galactic center with $m_{E} = m_{P} = 5$ GeV
and global ionized fraction $X_{E} = 0.1$.
The three lines corresponding to each distribution are 
calculated using best fit and $\pm \sigma$ deviations of the virial concentration parameter $C_{v}$ \cite{Catena:2009mf} 
which determines the inner slope of the Einasto profile. The vertical line at $r_{\odot} = 8.25 \,\mathrm{kpc}$ marks the local
 galactic position. While 1 $\sigma$ variations of the concentration parameter
  modifies these distributions by half an order of magnitude, their qualitative behavior is robust and 
the ionic density near the Sun's galactic position is generically suppressed by orders of magnitude relative to the global $X_{E}$.
 Similar corrections obtain unders $\pm 1\sigma$ variation in
 other CDM halo inputs (e.g. galactic virial mass --  local DM density); the local ionized
  fraction remains of order $X_{E}(r_{\odot})\sim 10^{-3}$. }
\label{fig:ionplot}
\end{center}
\end{figure}

For benchmark values of $m_{\rm ion} = 5\,\GeV$ and $\alpha_{D} = 0.1$, the condition in 
Eq.\,(\ref{eq:relax}) is trivially satisfied and the local ionized fraction becomes
\be
X_{E}(r_{\odot}) \equiv \frac{n_{\rm ion} (r_{\odot})}{n_{\rm ion} (r_{\odot}) + n_{CDM}(r_{\odot}) } \sim 10^{-3}~~.
\label{localfrac}
\ee
As the ion-ion scattering thermalizes, transferring heat from the core to the edge,
 the ions spread out away from each other to form an independent halo with farther 
reach than the atomic CDM distribution.
This dramatic local dilution opens up a new region of parameter space previously thought to be excluded by direct detection bounds.
In Figure \ref{fig:ionplot} we plot the radial profiles for both neutral (atomic) and ionized mass densities. 
 
 \par\bigskip\noindent
 Although the bullet cluster bounds allow global $X_{E}\lesssim 30\%$ \cite{Kaplan:2009de}, to be conservative, we will only consider 
 values around $10\%$ for the remainder of this paper. For larger global values, the assumptions of this section
are not satisfied and, furthermore, DM self-scattering constraints seem to rule out $X_E > 10\%$.  In any case, a dedicated numerical study is necessary to truly characterize the properties of the ionic halo. 
We also note that, unlike visible matter, our dark ions do not form a disk because the usual energy loss mechanisms
(e.g. cooling via bremsstrahlung and molecular de-excitation) are either suppressed or unavailable.

\subsection{\label{sec:dirdet} Direct Detection }

\noindent In this section we explore the $(M_{A},\,M_{\textrm{Atom}},\,E_{\textrm{hf}})$ parameter space in light of the positive signals at DAMA and CoGeNT, the constraints from XENON and CMDS, and recent preliminary results from CRESST \cite{CRESSTunpub}.  We will limit ourselves to portions of parameter space where the dark matter is primarily in atomic states, though this simplification still leaves four bound states to contend with: the chargitronia $(e,\tilde{e})\,\textrm{and}\,(p,\tilde{p})$ and the Hydrogen-like states $(e,p)\,\textrm{and}\,(\tilde{e},\tilde{p})$.  In order to predict count rates at the various experiments we need to know both their cross-sections for scattering from standard model nuclei and their relative cosmological abundances.  

\par\bigskip\noindent First, we consider scattering rates.  For a bound state of the form $(\mathcal{A},\mathcal{B})$ the interaction Hamiltonian which allows scattering off of standard model nuclei through a dark atomic hyperfine transition has the following form
\be
\hat{H}_{\textrm{int}} \sim Q_{\mathcal{A}}\frac{\vec{S}_{\mathcal{A}}\cdot\vec{q}}{\mu_{n\mathcal{A}}}\,F_{\mathcal{A}}\left(\frac{\mu_{\textrm{Atom}}}{m_{\mathcal{A}}}q\right)+Q_{\mathcal{B}}\frac{\vec{S}_{\mathcal{B}}\cdot\vec{q}}{\mu_{n\mathcal{B}}}\,F_{\mathcal{B}}\left(\frac{\mu_{\textrm{Atom}}}{m_{\mathcal{B}}}q\right),
\label{eq:hint}
\ee
where the $Q_{\mathcal{I}}$ are the axial charges of the atomic constituents, $\mu_{\textrm{Atom}}$ is the atomic reduced mass, the $\vec{S}_{\mathcal{I}}$ are the spin operators for the atomic constituents, $\vec{q}$ is the momentum transferred to the nucleus, the $m_{\mathcal{I}}$ are the masses of the atomic constituents, the $\mu_{n\mathcal{I}}$ are the reduced masses between nucleon and atomic constituents and the function $F_{\mathcal{I}}$ is the form factor for scattering off atomic constituent $\mathcal{I}$.  The scattering rate is then proportional to the matrix element of this Hamiltonian between initial and final dark atom - nucleus states.  In particular, the initial atomic state has total spin zero and the final atomic state is one of the three possible spin - 1 states.  For the chargitronia, the two terms in  Eq. \!(\ref{eq:hint})   are identical, so that interaction Hamiltonian is proportional to the \emph{total spin} of the chargitron.  For this reason, the chargitronium atoms do not scatter from ordinary nuclei at leading order in couplings.  In 
this regime, all the scattering rates have the same functional form, but  only a fraction of the total dark matter abundance able to scatter at direct detection experiments.  

\par\bigskip\noindent Given the above argument, it is important to understand the asymptotic value of $2Y_{ep}$, since the number of atoms able to scatter is proportional to this quantity.  Furthermore, the recombination rate for bound states is proportional to 
\be
\frac{\alpha^5 m^{3/2}_{\textrm{lite}}}{\sqrt{\mu_{\textrm{Atom}}}},
\ee
where $m_{\textrm{lite}}$ is the mass of the lightest atomic constituent; see our earlier work for details.  This indicates that $(p,\tilde{p})$ recombines most efficiently and $(e,\tilde{e})$ combines more efficiently than the Hydrogen-like states.  Note, however, that in the limit where all dark matter masses are equal, the recombination rates are equal.  If we consider case where this master recombination rate leaves very few ions around, then the final abundances of each of the four bound states will be one quarter of the total dark matter abundance and $2Y_{ep} = 1/2$.  For the remainder of this section we work in the equal mass limit and study the direct detection parameter space as a function of three parameters $M_{X}, E_{\textrm{hf}}\;\textrm{and}\;M_{\textrm{Atom}}$.

\par
\begin{figure}[!h]
\centering
\subfigure[\,$E_{\rm hf} = 5 \,\keV$ ]{\includegraphics[width=0.45\textwidth]{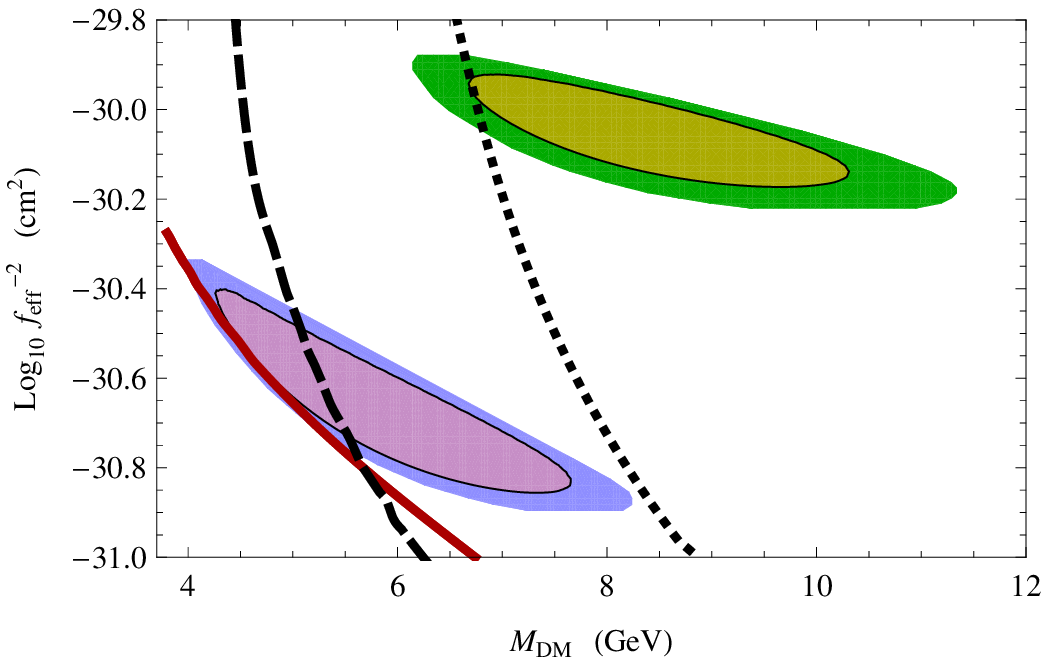}}
\subfigure[\,$E_{\rm hf} = 15 \,\keV$  ]{\includegraphics[width=0.45\textwidth]{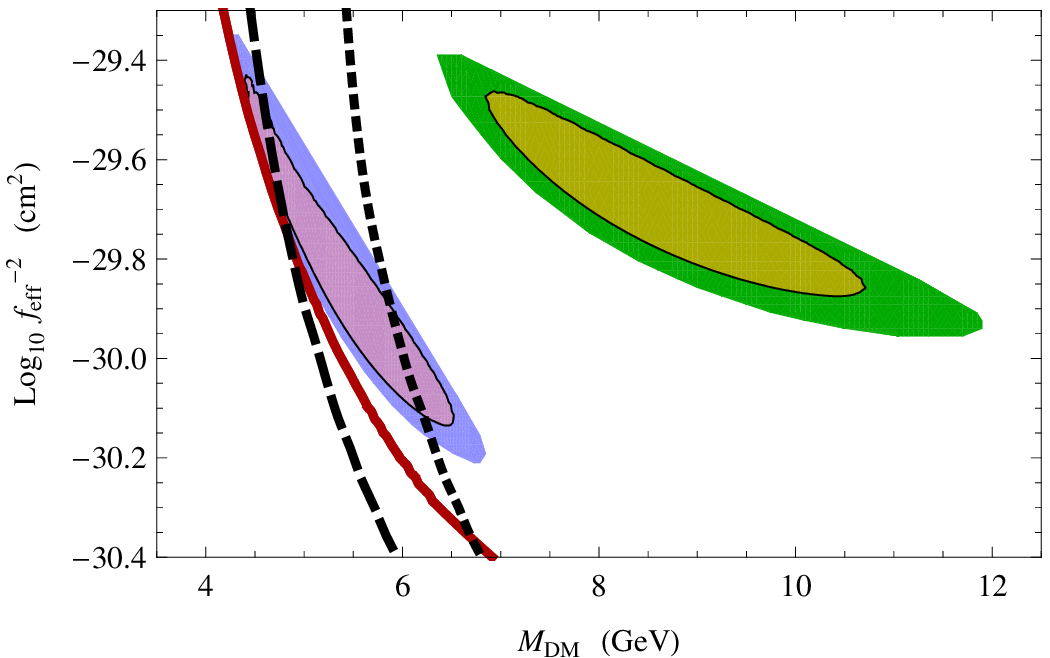}}
\caption{DAMA (yellow/green) and CoGeNT (purple/blue) 90\% and  95\% favored regions with CDMS-II Soudan exclusion lines (red, solid). 
 In (a) we also include the older XENON 10 bounds \cite{Angle:2007uj,Angle:2009xb} using the published low-recoil sensitivity (black, dashed) and 
 a modified efficiency which omits the lowest bin (black, dotted) to take into account the uncertainty in ${\cal L}_{eff}$. In (b) we use the most 
 recent XENON 10 release  \cite{Angle:2011th} which is more constraining for larger hyperfine splittings. Similar considerations result in two 
exclusion lines using the published low-threshold sensitivity (black, dashed) and a modified efficiency (black, dotted) with a 2 keV threshold. 
The \cog favored region is not constrained by XENON 100 because the low-energy threshold is above the characteristic 
nuclear recoil energies that explain \cog. Both plots assume a local dark matter density of $\rho_{dm} = 0.3\, \GeV/{\rm cm}^{3}$, however only 
$(ep)$ bound states scatter, so the effective density of scattering particles is $\rho_{dm}/2$. Following the discussion in 
Section 4.1, we neglect the effects of dark-ion scattering as their local density is highly suppressed.  }
\label{fig:m7}
\end{figure}

\bigskip\noindent In Figure \ref{fig:m7}, we find the 90\% and 95\% favored regions for DAMA and CoGeNT in the $f_{\textrm{eff}},\,M_{\textrm{DM}}$ parameter space for four different values of hyperfine splitting and with $m_{e}$ fixed to equal $m_p$; where $f_{\textrm{eff}}^4 \equiv M^4_{X}/(2(g_X\,\epsilon\,c_W)^2)$ controls the overall size of the scattering cross-section, $g_{X}$ is the \uone$_{X}$ coupling, and $c_{W}$ is the 
cosine of the weak mixing angle.  We find these regions by scanning over $\chi^2$ per degree of freedom, based on the spectra reported in \cite{Bernabei:2010ke} and \cite{Aalseth:2010vx} respectively.  In the DAMA case the $\chi^2$ is weighted by the reported uncertainties for each bin, whereas for CoGeNT we use Poisson statistics for the uncertainties.  Figure \ref{fig:m7} also includes constraint lines for XENON10 \cite{Angle:2009xb} and the low-threshold re-analysis of CDMS Ge \cite{Ahmed:2010wy} where we have also used Poisson statistics to define the error bars.  To account for the controversy over the low-threshold behavior of $\mathcal{L}_{eff}$ bin at XENON, we plot a modified exclusion line which omits the 2 - 5 keV recoil bin entirely.  Any other treatment of the low-threshold behavior of XENON's detector  interpolates between these two contours.  The CDMS exclusion is calculated via a $\chi^2$ by taking the 95\% confidence limit of the spectrum reported in \cite{Ahmed:2010wy} and weighting the $\chi^2$ with Poisson uncertainties.  We find that this method adequately reproduces the ``vanilla'' WIMP exclusion lines reported by CDMS.  

\par\bigskip\noindent A few comments are in order.  First, we see that while increasing the hyperfine splitting moves the DAMA and \cog regions closer to one another -- as one would expect since Germanium is a heavier nucleus than Sodium  -- there is no overlap between the two.  As the hyperfine splitting is pushed to even higher values the CoGeNT allowed region becomes a very narrow, nearly vertical strip around 6 GeV. 
We note that variations in the galactic CDM Halo -- especially the escape velocity -- as well as known uncertainties in the DAMA quenching can improve agreement between DAMA and CoGeNT   \cite{Hooper:2010uy}.  The regions plotted above are conservative in the sense that they do not take advantage of these variations.
 Second, note that the aDM parameter space favored by DAMA is completely ruled out by the most recent CDMS analysis and the more constraining XENON exclusion, while the less aggressive treatment of XENON's low-threshold behavior does leave some parameter space for DAMA\footnote{We also point out that the tension between DAMA and CoGeNT is not alleviated by ignoring the shape of the DAMA spectrum and considering only the net count rate.}.  Third, note that increasing the hyperfine splitting does not have much of an effect on the allowed region for CoGeNT.  This is reasonable, given that CDMS puts the tightest constraints on aDM and both CoGeNT and CDMS both look for Ge recoils.  Dark atoms are not ruled out by the low-threshold results of CDMS or XENON, while light WIMPS apparently are, because the aDM recoil spectrum goes to zero linearly at low energies.  In contrast, WIMP scattering is exponentially more likely at low recoil.

\par\bigskip\noindent Finally, there is the matter of CRESST.  Since the CRESST detector is made of Calcium - Tungstate (CaWO$_4$) crystals, and the Oxygen/Tungsten recoils bands are distinguishable, CRESST is able to contemporaneously search for light DM scattering and heavy DM scattering, respectively.  \emph{Preliminary} results suggest that with $\mathcal{O}(550)$\,kg-days of exposure CRESST sees roughly 23 events \emph{in the Oxygen band}\,\cite{CRESSTunpub}.  We find that the regions preferred by CoGeNT for $E_{\textrm{hf}} = 5,\,15$\,keV are consistent at the 90\% confidence level, with the count rate in Oxygen at CRESST.  We find that, generically, the DAMA preferred region predicts a count rate at CRESST which is about four times too large.
%
%\begin{figure}[!h]
%\centering
%\subfigure[\,$E_{\rm hf} = 1 \,\keV$  ]{\includegraphics[width=0.45\textwidth]{CRESST_hf1.eps} \label{fig:CRESST1}}
%\subfigure[\,$E_{\rm hf} = 10 \,\keV$  ]{\includegraphics[width=0.45\textwidth]{CRESST_hf10.eps} \label{fig:CRESST2}}
%\subfigure[\,$E_{\rm hf} = 15 \,\keV$  ]{\includegraphics[width=0.45\textwidth]{CRESST_hf15.eps} \label{fig:CRESST3}}
%\caption{DAMA (red - yellow) and CoGeNT (purple - tan) 90\% and  95\% favored regions plotted with contours of constant event rate for CRESST's preliminary Oxygen results for $E_{\textrm{hf}} = 1,\,10$\,keV. We have also plotted the CoGeNT preferred region with CRESST counts for $E_{\textrm{hf}} = 15$\,keV.}
%\label{default}
%\end{figure}

%
%  
%\begin{figure}[!h]
%\begin{center}%
%\hspace*{-2.5cm}
%\includegraphics{mainfigPositroniumHf1.eps}
%\caption{Plot of the CoGeNT and DAMA favored regions at 90 and 95 \% confidence contours with $E_{hf} = 5 $ KeV and $M_{e} = 2$ \GeV. The red line is the CDMS exclusion at 95\% confidence. }
%\label{direct}
%\end{center}
%\end{figure}

\section{\label{sec:disc} Discussion }
\noindent In this article we have studied the rich cosmology and parameter space of atomically bound dark matter.  The abundance of dark atoms can be tied to the baryon asymmetry in which the decays of heavy sterile neutrinos
 generate both dark and visible sector abundances. For natural couplings to heavy neutrinos, both sectors acquire equal number densities, so the dark sector mass scale must be $\mathcal{O}(5\,\textrm{GeV})$
 to reproduce the observed DM abundance.  Since the gauge field that binds the dark atoms must be embedded in a non-Abelian group to avoid a Landau pole below the Planck scale, the dark matter
  is divided into four atomic species whose asymptotic abundances are very sensitive to the dark fine structure constant and the mass of each binding combination.  The ionic species generically interact rapidly enough to maintain kinetic equilibrium and thereby form a separate, more diffuse halo than that of the cold atoms.  

\par\bigskip\noindent Our analysis has emphasized the limit where all atomic constituents have equal masses. By symmetry, the atomic species
 in this limit comprise equally abundant populations of  ``chargitronium.''     Because the dark atoms are 
light compared to the weak scale, the most significant constraints on aDM come from the low-threshold reanalyses at CDMS and XENON10/100.  
While there is significant tension between DAMA and CoGeNT, the parameter space favored by the CoGeNT signal -- and allowed by null results -- predicts a 
large signal at CRESST of the right order to explain the excess reported in preliminary results.

\par\bigskip\noindent  There are a number of directions for further study.  The cosmology of aDM is intricate and a full numerical study of the parameter space for both the asymmetry and recombination would be interesting.  Furthermore, while it is clear that kinetic equilibrium will lead to a distinct ionic halo, the details of the aDM phase space distribution can only be determined through numerical simulations, which require knowledge of the initial power spectrum.  It would also be interesting to consider the observational consequences of the ionic halo; for example, in principle there could be  long range dipole-dipole interactions between galactic halos.  For simplicity, the model has an exact parity that prevents dark atom decay.  It would be interesting to consider soft violations of this parity and the potentially observable consequences.  Finally, we have only studied the direct detection parameter space only in the case of a fully degenerate dark sector.  Since both the abundance of atoms and shape of the direct detection spectrum are sensitive to the masses of the atomic constituents, the parameter space for more generic combinations is difficult to map.  The possibility of better agreement between the various positive signals and null results makes a more thorough study valuable.  

\acknowledgments We thank Bogdan Dobrescu, Patrick Fox, Roni Harnik, Joachim Kopp, Graham Kribs,  
and Raman Sundrum,  for helpful conversations.  CMW was partially supported by the Houghton College Summer Research Institute.  GZK is supported by a Fermilab Fellowship in Theoretical Physics.
Fermilab is operated by Fermi Research Alliance, LLC, under Contract DE-AC02-07CH11359
with the US Department of Energy. This work is supported by the U.S. Department of Energy under cooperative research agreement Contract Number DE-FG02-05ER41360.

%\bibliography{aDMbib.bib}{}
%\bibliographystyle{apsrev4-1}

\end{document}